\documentclass[aps,prb,twocolumn,superscriptaddress]{revtex4}

\usepackage{times}
\usepackage{amsmath,bm,amsfonts}
\usepackage{dcolumn} 
\usepackage{graphicx}
\usepackage{latexsym}

\usepackage{color}

\begin{document}

\title{Anisotropic Exchange Interactions of Spin-Orbit-Integrated
  States in Sr$_{2}$IrO$_{4}$}

\author{Hosub \surname{Jin}}

\affiliation{Department of Physics and Astronomy and Center for Strongly
  Correlated Materials Research, Seoul National University, Seoul
  151-747, Korea}

\author{Hogyun \surname{Jeong}}

\affiliation{Computational Science and Technology Interdisciplinary
  Program, Seoul National University, Seoul 151-747, Korea}

\author{Taisuke \surname{Ozaki}}

\affiliation{Research Center for Integrated Science, Japan Advanced
  Institute of Science and Technology, Nomi, Ishikawa 923-1292, Japan}

\author{Jaejun \surname{Yu}}
\email[Corresponding author. Electronic address: ]{jyu@snu.ac.kr}
\affiliation{Department of Physics and Astronomy and Center for Strongly
  Correlated Materials Research, Seoul National University, Seoul 151-747,
  Korea}
\affiliation{Center for Theoretical Physics, Seoul National University,
  Seoul 151-747, Korea}

\date{\today}

\begin{abstract}
  We present a microscopic model for the anisotropic exchange interactions
  in Sr$_{2}$IrO$_{4}$. A direct construction of Wannier functions from
  first-principles calculations proves the $j_{\mathrm{eff}}$=1/2
  character of the spin-orbit integrated states at the Fermi level. An
  effective $j_{\mathrm{eff}}$-spin Hamiltonian explains the observed weak
  ferromagnetism and anisotropy of antiferromagnetically ordered magnetic
  state, which arise naturally from the $j_{\mathrm{eff}}$=1/2 state with
  a rotation of IrO$_{6}$ octahedra.  It is suggested that
  Sr$_{2}$IrO$_{4}$ is a unique class of materials with effective exchange
  interactions in the spin-orbital Hilbert space.
\end{abstract}

\pacs{75.30.Et, 71.70.Ej, 71.20.-b, 75.30.Gw}

\maketitle

\section{Introduction}
\label{sec:introduction}

Many of transition metal oxides (TMOs) are antiferromagnetic insulators.
The simplest model for such Mott insulators is the Hubbard model
Hamiltonian,\cite{hubbard} which gives rise to an effective exchange term
called superexchange interaction at half-filling. When orbital degrees of
freedom are involved, a variety of exchange interactions can occur for a
given ionic configuration with different crystal structures. In the case
of colossal magneto-resistance manganese oxides, for instance, the
superexchange interaction with orbital degeneracy determines complex spin
and orbital orderings and, when doped, degenerate $e_{g}$ orbitals coupled
to the lattice via Jahn-Teller interactions become an essential part of
the double exchange physics\cite{maekawa}. Sometimes the orbital degrees
of freedom via spin-orbit (SO) coupling are responsible for the magnetic
anisotropy bound to the crystal environment. When there exists an orbital
degeneracy, SO coupling may become a dominant term so that the effective
Hamiltonian should involve the full spin-orbital Hilbert space where the
ground state must comply with the intersite spin and orbital
correlations\cite{jackeli}. A possible dynamic interference between the
spin and orbital space was suggested in vanadates\cite{zhou:156401}. There
was a report of a large spin-orbital fluctuations in Mott insulators with
$t_{2g}$ orbital degeneracy as a manifestation of quantum entanglement of
spin and orbital variables \cite{oles:147205}.

Recently we have shown that the electron correlation effect combined with
strong spin-orbit (SO) interactions is responsible for the observed
insulating behavior of 5$d$ TMO Sr$_{2}$IrO$_{4}$ \cite{kim:076402}. While
SO coupling has been considered as a minor perturbation in the description
of magnetism \cite{yosida}, the amount of SO interactions in 5$d$ elements
including Ir, for example, is an order of magnitude larger than in the
3$d$ TMO system \cite{PhysRevB.13.2433}.
Thus the SO coupling is expected to play a significant role in the
electronic and magnetic properties of 5$d$ TMO systems. Indeed the
manifestation of a novel $j_{\mathrm{eff}}$=1/2 Mott ground state in
Sr$_{2}$IrO$_{4}$ was revealed by angle resolved photoemission
spectroscopy, optical conductivity, x-ray absorption spectroscopy
measurements, and first-principles electronic structure calculations
\cite{kim:076402}.  Further investigations of the electronic structures of
the Sr$_{n+1}$Ir$_{n}$O$_{3n+1}$ ($n=$ 1, 2, and $\infty$) series
demonstrated a Mott insulator-metal transition with a change of bandwidth
as $n$ increases\cite{moon:226402}. The ground state of 5$d$ TMO
Sr$_{2}$IrO$_{4}$ is a Mott insulator in the strong spin-orbit coupling
limit. In addition, Sr$_{2}$IrO$_{4}$ exhibits unusual weak ferromagnetism
with reduced Ir magnetic
moments\cite{PhysRevB.49.9198,PhysRevB.52.9143,PhysRevB.57.R11039,Kini.2006,moon:113104}.
To understand such unusual magnetic properties of Sr$_{2}$IrO$_{4}$, it is
necessary to take account of the spin-orbit integrated
$j_{\mathrm{eff}}$=1/2 state.

In this paper, we introduce a prototype model of spin-orbit-integrated
magnetism realized in Sr$_{2}$IrO$_{4}$. From a tight-binding analysis
based on first-principles calculations, we show that the
$j_{\mathrm{eff}}$=1/2 character of the spin-orbit-integrated state
remains robust even in the presence of on-site Coulomb interactions. A
direct construction of Wannier functions from first-principles
calculations proves the $j_{\mathrm{eff}}$=1/2 character at the Fermi
level. An effective exchange Hamiltonian with not $S$=1/2 but
$j_{\mathrm{eff}}$=1/2 is obtained starting from a $j_{\mathrm{eff}}$=1/2
Hubbard model. The origin of anisotropic magnetic exchange interactions
are discussed in connection with an extraordinary character of the ground
state. The presence of spin-orbit-integrated state with strong SO
interactions in Sr$_{2}$IrO$_{4}$ can make 5$d$ Ir-oxides a unique class
of materials for the study of effective exchange interactions in the full
spin-orbital Hilbert space.

\section{Spin-Oribt-Integrated Electronic States}
\label{sec:spin-oribt-integr}

\subsection{LDA+SO+$U$ Band Structure}
\label{sec:lda+s-band-struct}

Since both on-site Coulomb interactions ($U$) and SO couplings are
expected to be important in the description of Ir 5$d$ states, we examined
the effect of on-site $U$ and SO couplings separately and simultaneously
on the electronic structure of Sr$_2$IrO$_4$. To identify the role of each
term and the interplay between them, we carried out
density-functional-theory (DFT) calculations within the local-density
approximation (LDA), LDA including the SO coupling (LDA+SO), and LDA+$U$
including the SO coupling (LDA+SO+$U$), respectively. We calculated total
energies and electronic band structures of Sr$_{2}$IrO$_{4}$ for the
structural parameters as obtained from the neutron powder diffraction data
at 10 K \cite{huang1994}, which has a K$_2$NiF$_4$-type layered perovskite
structure with the symmetry of the space group $I4_1/acd$ reduced from
$I4/mmm$, where IrO$_6$ octahedra are rotated by about 11$^{\circ}$ around
the $c$-axis of the unit cell. For the calculations, we used the DFT code,
OpenMX \cite{openmx}, based on the
linear-combination-of-pseudo-atomic-orbitals method
\cite{PhysRevB.67.155108}, where both the LDA+$U$ method \cite{han:045110}
and the SO couplings were included via a relativistic $j$-dependent
pseudo-potential scheme in the non-collinear DFT formalism
\cite{MacDonald,PhysRevB.26.4199,PhysRevB.64.073106}.
Double valence and single polarization orbitals were used as a basis set,
which were generated by a confinement potential scheme with cutoff radii of
8.0, 7.0 and 5.0 a.u. for Sr, Ir, and O atoms respectively.  We used a
(6$\times$6$\times$4) \textbf{k}-point grid for the k-space integration.

\begin{figure}[t]
 \centering\includegraphics[width=8cm]{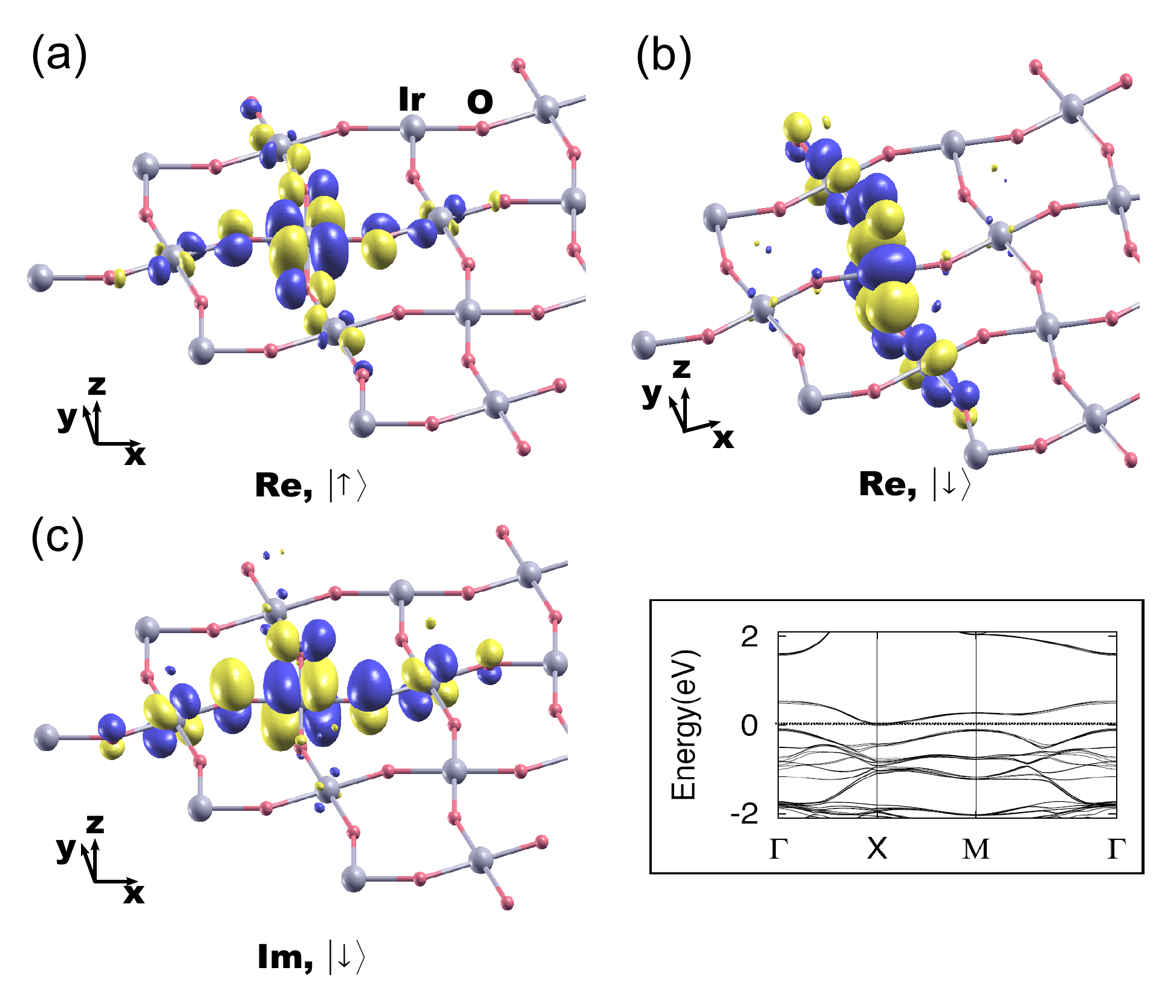}
 \caption{(Color online) Calculated Wannier functions of the
   $|j_{\mathrm{eff}}=1/2,+1/2 \rangle$ state: (a) the real part of
   up-spin $|\uparrow\rangle$ component, (b) the real part of the
   down-spin $|\downarrow\rangle$ component, and (c) the imaginary part of
   the down-spin $|\downarrow\rangle$ component. Blue (dark gray) and yellow (light gray)  colors in the Wannier  
   function represent negative and positive values respectively. The inset shows the
   LDA+SO+$U$ band structure near $E_{\mathrm{F}}$=0 eV, emphasizing the
   ``upper'' and ``lower'' Hubbard band of the $|j_{\mathrm{eff}}=1/2,m
   \rangle$ band above and below $E_{\mathrm{F}}$
   respectively.} \label{fig:1}
\end{figure}

Calculated LDA, LDA+SO, and LDA+SO+$U$ band structures were presented in
the previous work\cite{kim:076402} where the results of the LDA+SO+$U$
band structures of Sr$_{2}$IrO$_{4}$ are well compared with those of angle
resolved photoemission spectroscopy experiments. The LDA bands near the
Fermi level ($E_{\mathrm{F}}$), as shown in Fig.~2(a) of
Ref.~[\onlinecite{kim:076402}], are almost identical to those of
Sr$_{2}$RhO$_{4}$, \cite{kim:106401} which can be expected from the same
$d^{5}$ configuration of Rh$^{4+}$ and Ir$^{4+}$ and the same structural
distortions, i.e., the rotations of RhO$_{6}$ and IrO$_{6}$ octahedra. The
hybridization of $d_{xy}$ and $d_{x^{2}-y^{2}}$ due to the rotation of
IrO$_{6}$ octahedra pushes the $d_{xy}$ band below $E_{\mathrm{F}}$,
similarly to the case of Sr$_{2}$RhO$_{4}$. Indeed the LDA Fermi surface
of Sr$_{2}$IrO$_{4}$ was found to be basically the same as that of
Sr$_{2}$RhO$_{4}$ \cite{kim:106401}.

In the LDA band structure, the contribution of $d_{xy}$ components above
$E_{\mathrm{F}}$ are strongly suppressed relative to those of $d_{yz}$ and
$d_{zx}$ states, whereas the Ir 5$d$ bands ranging from $-2.5$ eV to 0.5
eV are still dominated by the $t_{2g}$ orbitals with a small admixture of
$d_{x^{2}-y^{2}}$. On the other hand, however, when the SO coupling is
included, a significant change of the wave function character occurs so
that all three $t_{2g}$ orbital components are almost equally distributed
in the LDA+SO band structure. This change arises from the SO interactions
acting on the $t_{2g}$ manifold, which mixes up the $d_{xy}$, $d_{yz}$,
and $d_{zx}$ orbitals. This qualitative change of the wave function
character is an essence of the SO coupling action, which is related to the
novel nature of the SO-integrated insulating ground state.


As shown in the inset of Fig.~\ref{fig:1}, an effective $U$ = 2 eV opens
up a gap in the LDA+SO+$U$ band structure and gives rise to the
non-dispersive and parallel features of ``upper'' and ``lower'' Hubbard
bands of the SO-integrated states, which are in excellent agreement with
experimental observations \cite{kim:076402}. It is remarkable to obtain an
insulating ground state for the intermediate value of $U$, which is
smaller than the band width of the $t_{2g}$ manifold and the conventional
$U$ values of 3$d$ TMOs. On the other hand, however, when we performed
LDA+$U$ calculations without the SO coupling, the on-site Coulomb
interaction became \textit{ineffective} due to the three-fold degeneracy
of the Ir $t_{2g}$ manifold crossing $E_{\mathrm{F}}$ \cite{footnote}.
Unless the degeneracy is broken, each band remains partially filled being
far from the Mott instability. In our non-collinear DFT calculations, the
SO coupling terms were solved in a completely non-perturbative way,
whereas the Coulomb correlation effect were treated via the LDA+$U$
method.

\begin{table}
  \caption{Coefficients of the Wannier functions illustrated in Fig.~\ref{fig:1}:
    Only the coefficients from the center Ir are listed.}\label{tab:1}
  \begin{center}
    \begin{ruledtabular}
    \begin{tabular}{llrrrr}
      & & \multicolumn{2}{c}{up-spin} & \multicolumn{2}{c}{down-spin} \\ \cline{3-6}
      & & Re\phantom{aa}  & Im\phantom{aa} & Re\phantom{aa} & Im\phantom{aa} \\ \hline
      Ir  &  $d_{z^2}$ &   0.00009  &   0.00009  &   0.00002  &   0.00001 \\
      &  $d_{x^2-y^2}$ &  -0.09044  &  -0.01212  &  -0.00024  &   0.00016 \\
      &  $d_{xy}$      &   0.32738  &   0.00000  &   0.00115  &  -0.00009 \\
      &  $d_{yz}$      &  -0.00203  &  -0.00001  &   0.44105  &   0.05527 \\
      &  $d_{zx}$      &  -0.00018  &   0.00183  &  -0.05450  &   0.44212 \\
    \end{tabular}
    \end{ruledtabular}
  \end{center}
\end{table}

In order to examine the nature of the SO-integrated state, we constructed
Wannier functions which can identify the orbital shape and bonding
character of the ``upper'' and ``lower'' Hubbard bands of the
SO-integrated states, as shown in the inset of Fig.~\ref{fig:1}. The
Wannier functions were calculated for the $t_{2g}$ manifold by employing
the projection scheme\cite{anisimov:125119}. The Wannier function illustrated in
Fig.~\ref{fig:1} corresponds to the single band above $E_{\mathrm{F}}$ of
LDA+SO+$U$ bands. As listed in Table~\ref{tab:1}, the overall shape of the
calculated Wannier function of Fig.~\ref{fig:1} matches closely to the
ideal $j_{\mathrm{eff}}$=1/2 state: \begin{equation}
  \label{eq:6}
  |j_{\mathrm{eff}}=\frac{1}{2},\pm \frac{1}{2}\rangle =
  \mp\frac{1}{\sqrt{3}} \left[ |d_{xy}\rangle|\pm\rangle \pm
    \left( |d_{yz}\rangle\pm i |d_{zx}\rangle \right)|\mp\rangle\right] ,
\end{equation}
where $|\pm\rangle$ represent for the up-spin $|\uparrow\rangle$ and
down-spin $|\downarrow\rangle$ states respectively. The agreement of its
orbital components and their relative phases between the ideal state and
the calculated Wannier function is another proof of the SO-integrated
$j_{\mathrm{eff}}$=1/2 state. Here, for the sake of simplicity in the
presentation, we chose a self-consistent solution with the spin
quantization axis parallel to the $z$-axis.

\subsection{Tight-binding model}
\label{sec:tight-binding-model}

The physics of the LDA+SO+$U$ results can be captured by a multi-band
Hubbard model for the $t_{2g}$ bands including the SO coupling term. The
tight-binding (TB) bands for the $t_{2g}$ manifold can be described by
\begin{equation}
  \label{eq:7}
  \mathcal{H}_{0} = \sum_{\langle ij\rangle\alpha\beta\sigma}
  t^{\alpha\beta}_{ij} c^{\dagger}_{i\alpha\sigma}c_{j\beta\sigma} +
  \sum_{i,a=d_{xy}} \Delta_{t}c^{\dagger}_{ia\sigma}c_{ja\sigma}
  +\lambda_{\mathrm{SO}} \sum_{i} \mathbf{L}_{i}\cdot\mathbf{S}_{i} ,
\end{equation}
where $\langle ij\rangle$ runs over the nearest neighbor pairs of sites
$i$ and $j$ in the two-dimensional square lattice consisting of Ir ions,
$\alpha$ and $\beta$ are indices for $t_{2g}$ orbitals, i.e., $\{ d_{xy},
d_{yz}, d_{zx}\}$, $t^{\alpha\beta}_{ij}$ a hopping integral between
$|i\alpha\rangle$ and $|j\beta\rangle$, $\Delta_{t}$ a tetragonal crystal
field splitting, i.e., an on-site energy difference of the $d_{xy}$
orbital relative to $d_{yz}$ and $d_{zx}$, and $\lambda_{\mathrm{SO}}$ the
SO coupling parameter. In a simple square lattice of Ir ions,
$t^{\alpha\beta}_{ij}$ becomes a non-zero constant $t_{0}$ only for
$(\alpha,\beta)=(d_{xy},d_{xy})=(d_{zx},d_{zx})$ with
$j=i+\hat{\mathbf{x}}$ and so on.

Starting from a set of $\{ \langle n_{i\tau\alpha\sigma}\rangle\}$ as
mean-field parameters, we could obtain a self-consistent mean-field Hamiltonian
within the $t_{2g}$ subspace by $\mathcal{H}_{t_{2g}} = \sum_{\mathbf{k}}
C^{\dagger}_{\mathbf{k}} \widehat{\mathcal{T}}(\mathbf{k}) C_{\mathbf{k}}$ where
$C_{\mathbf{k}}$ has 12 components of $\{ c_{\mathbf{k}\tau\alpha\sigma} |
\tau=A,B; \alpha=d_{xy},d_{yz},d_{zx}; \sigma = \uparrow,
\downarrow\}$. Here the site indices $\tau=A,B$ are for the two
inequivalent Ir sites. By choosing the basis in order of
$(c_{A d_{xy}\uparrow},c_{A d_{yz}\downarrow},c_{A d_{zx}\uparrow},
[A\rightarrow B], [\uparrow\leftrightarrow\downarrow])$, we can find a
block-diagonal 12$\times$12 $\widehat{\mathcal{T}}(\mathbf{k})$ matrix:
\begin{equation}
  \label{eq:2}
  \widehat{\mathcal{T}}(\mathbf{k}) = \left( \begin{array}{c c | c c}
      \mathbf{D}_{\mathrm{I}}^A & \mathbf{O}(\mathbf{k})& 0 & 0 \\
      \mathbf{O}^{\dagger}(\mathbf{k}) & \mathbf{D}_{\mathrm{I}}^B & 0 & 0 \\ \hline
      0 & 0 & \mathbf{D}_{\mathrm{II}}^A & \mathbf{O}(\mathbf{k}) \\
      0 & 0 & \mathbf{O}^{\dagger}(\mathbf{k}) & \mathbf{D}_{\mathrm{II}}^B
\end{array} \right) .
\end{equation}
Here the hopping integrals contribute to
$\mathbf{O}(\mathbf{k})$:
\begin{equation}
  \label{eq:3}
  \mathbf{O}(\mathbf{k}) = e^{-i \frac{k_x+k_y}{2}}\left( \begin{array}{c c c}
      -4t_{0} \gamma_{1\mathbf{k}}  & 0 & 0 \\
      0 & -2t_{0} \gamma_{2\mathbf{k}} & 0 \\
      0 & 0 & -2t_{0} \gamma_{3\mathbf{k}}
    \end{array} \right),
\end{equation}
where the non-zero hopping terms of
$t_{0}=t_{d_{xy}}=t_{d_{yz}}=t_{d_{zx}}$ lead to the dispersions
$\gamma_{1 \mathbf{k}}=\cos{\frac{k_x}{2}} \cos{\frac{k_y}{2}}$,
$\gamma_{2 \mathbf{k}}=\cos{\frac{k_x+k_y}{2}}$, and $\gamma_{3
  \mathbf{k}}=\cos{\frac{k_x-k_y}{2}}$ for $d_{xy}$, $d_{yz}$, and
$d_{zx}$ bands, respectively.  The on-site Coulomb interaction $U$ and the
SO coupling $\lambda_{\mathrm{SO}}$ contribute to the diagonal term:
\begin{equation}
  \label{eq:5}
  \mathbf{D}_{\mathrm{I}}^{\tau} = \left( \begin{array}{c c c}
      \Delta_{t} + e_{1}\gamma_{1\mathbf{k}}^{2} -  U\bar{n}_{\tau d_{xy}
        \uparrow} & \lambda_{\mathrm{SO}}/2 & -i \lambda_{\mathrm{SO}}/2 \\
      \lambda_{\mathrm{SO}}/2 & -U\bar{n}_{\tau d_{yz} \downarrow} & -i
      \lambda_{\mathrm{SO}}/2 \\
      \imath \lambda_{\mathrm{SO}}/2 & i \lambda_{\mathrm{SO}}/2 &
      -U\bar{n}_{\tau d_{zx} \downarrow} \\
\end{array} \right),
\end{equation}
and $\mathbf{D}_{\mathrm{II}}^{\tau}$ is a time-reversal partner of
$\mathbf{D}_{\mathrm{I}}^{\tau}$.

Despite of a large cubic crystal field splitting due to
$\Delta_{c}\approx$ 5 eV between $t_{2g}$ and $e_{g}$, there is a
significant hybridization of $d_{xy}$ and $d_{x^{2}-y^{2}}$ due to the
rotation of IrO$_{6}$ octahedra. In order to describe both LDA and LDA+SO
band structures properly, the contribution of the $d_{x^{2}-y^{2}}$
admixture is necessary to be included as a $\mathbf{k}$-dependent energy
$\Delta \varepsilon_{\mathbf{k} d_{xy}}$ for the $d_{xy}$ band: $\Delta
\varepsilon_{\mathbf{k} d_{xy}} = e_{1} \gamma^{2}_{1\mathbf{k}}$. The
best fit to the LDA bands was obtained by a set of parameters:
$\Delta_{t}$= 0.15eV, $t_0$=0.35eV, and $e_{1} = -1.5$eV.

\begin{figure}[t]
 \centering\includegraphics[width=8cm]{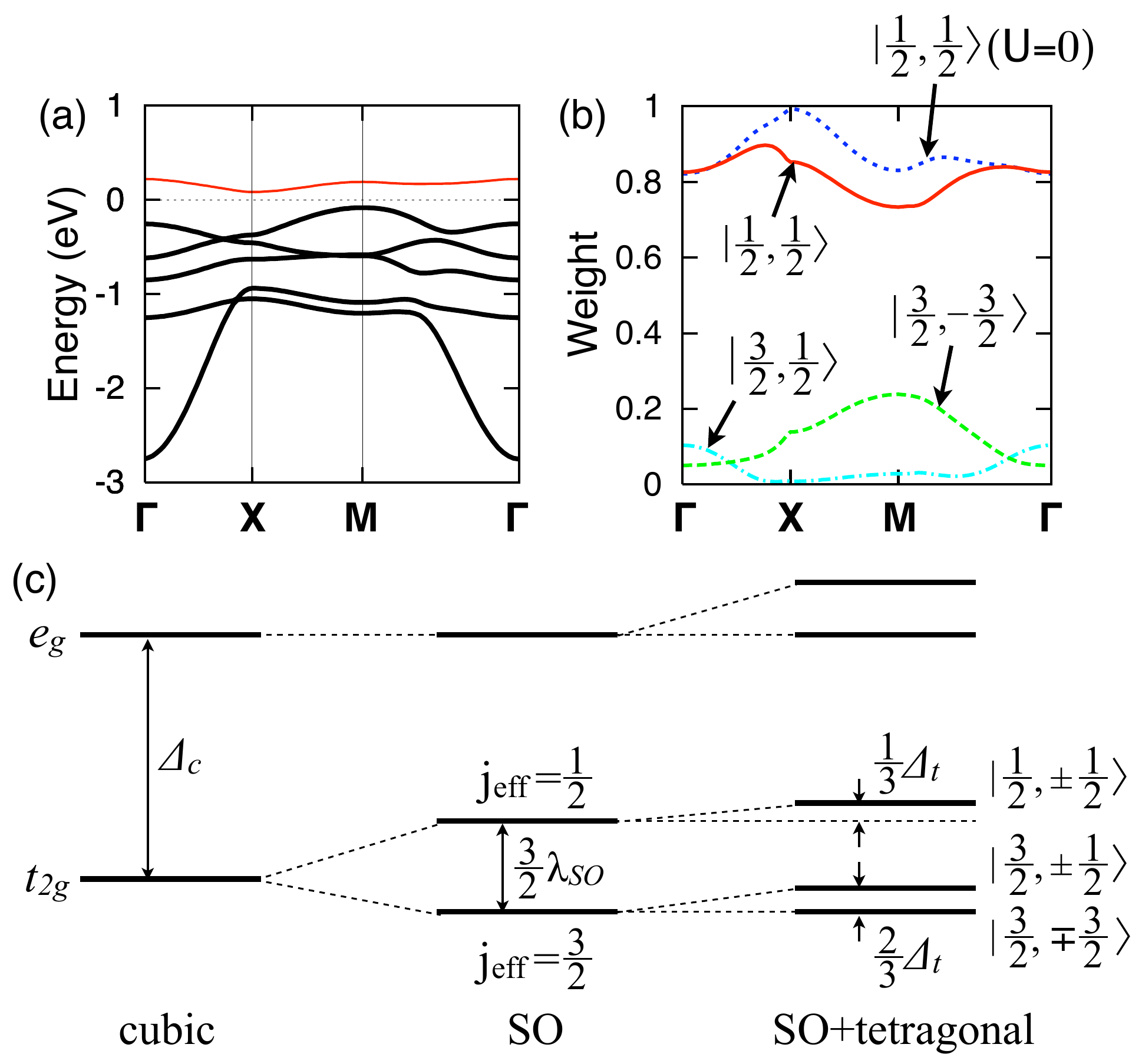}
 \caption{(Color online) (a) Tight-binding band structure with
   ($\lambda_{\mathrm{SO}}$, $U$)=(0.4 eV, 2.0 eV), which is well compared
   with the LDA+SO+$U$ band structure of Fig.~\ref{fig:1}(c), (b) the
   decomposition of the ``upper'' Hubbard band wavefunction (marked by a
   thin (red) solid line in (a)) into the $\{|j_{\mathrm{eff}},m_j
   \rangle\}$ basis, where the dotted line represents a
   $|\frac{1}{2},\frac{1}{2}\rangle$ component with $U$=0, and (c) a
   schematic energy diagram of the $t_{2g}$ manifold in the atomic limit.
   Due to the large crystal field splitting by $\Delta_c$, the $t_{2g}$
   levels can be mapped into the effective $l_{\mathrm{eff}}=1$ states
   where the tetragonal field splitting by $\Delta_{t}$ is relatively
   insignificant.} \label{fig:2}
\end{figure}

The solutions of our TB model including both SO coupling
$\lambda_{\mathrm{SO}}$ and on-site Coulomb interaction $U$ with different
sets of parameters ($\lambda_{\mathrm{SO}}$, $U$)=(0, 0), (0.4 eV, 0) and
(0.4 eV, 2.0 eV) reproduce well the $t_{2g}$ manifold of the LDA, LDA+SO,
and LDA+SO+$U$ bands, respectively. The self-consistent solution for
($\lambda_{\mathrm{SO}}$, $U$)=(0.4 eV, 2.0 eV) is shown in
Fig.~\ref{fig:2}(a), which corresponds to the LDA+SO+$U$ bands of the
inset of Fig.~\ref{fig:1}.

In addition to the large crystal field splitting between $e_{g}$ and
$t_{2g}$, the $t_{2g}$ manifold splits further into doubly degenerate
$j_{\mathrm{eff}}$=1/2 and quadruply degenerate $j_{\mathrm{eff}}$=3/2
states due to SO coupling. The small tetragonal crystal field does not
affect this configuration. A schematic energy level diagram shown in
Fig.~\ref{fig:2}(c) was confirmed by the LDA and LDA+SO energy levels at
the $X$ point,\cite{kim:076402} where the off-diagonal hopping matrix
$\mathbf{O}(\mathbf{k})$ in the TB model becomes zero. Even though the
non-zero hopping terms away from $X$ point may disturb the atomic picture,
the SO coupling retains the anti-crossing between those levels which
transform according to the same irreducible representation
\cite{winkler03}. Consequently, the effective band-width of the
half-filled $j_{\mathrm{eff}}$=1/2 band becomes smaller than the modest
value of on-site $U$. The decomposition of the ``upper'' Hubbard band
wavefunction into the $\{|j_{\mathrm{eff}},m_j \rangle\}$ basis clearly
demonstrates the robustness of its $j_{\mathrm{eff}}$=1/2 character as
shown in Fig.~\ref{fig:2}(b), whereas the $j_{\mathrm{eff}}$=1/2 weight
for $U=2$ eV is slightly reduced from that of $U=0$. Therefore it is
reasonable to consider an effective Hamiltonian based on the
$j_{\mathrm{eff}}$=1/2 single-band Hubbard model instead of the
conventional $S$=1/2 model: \begin{equation}
  \label{eq:8}
  \mathcal{H} = \sum_{\langle ij\rangle m m'} \bar{t}^{ij}_{mm'}
  d^{\dagger}_{im}d_{jm'} + \bar{U} \sum_{i} n_{d i +1/2}n_{d i -1/2} ,
\end{equation}
where $d_{im}$ represents for the $|j_{\mathrm{eff}}=1/2,m \rangle$ state
at the site $i$ with $m, m'=\pm 1/2$ and $n_{d i m} =
d^{\dagger}_{im}d_{im}$. $\bar{t}^{ij}_{mm'}$ and $\bar{U}$ are effective
hopping and on-site interaction parameters respectively.

\section{Anisotropic Exchange Interactions}
\label{sec:anis-exch-inter}

\subsection{Effective Exchange Hamiltonian}
\label{sec:anis-exch-inter-1}

The $j_{\mathrm{eff}}$=1/2 single-band Hubbard model has an interesting
feature in $\bar{t}^{ij}_{mm'}$, which originates from a peculiar nature
of the spin-orbit integrated state in Sr$_{2}$IrO$_{4}$. In the strong SO
coupling limit, the orbital wavefunctions of the $|j_{\mathrm{eff}}=1/2,m
\rangle$ state of Eq.~\eqref{eq:6} consists of the cubic harmonics with
respect to the local coordinate axes. The rotation of the IrO$_{6}$
octahedron results in a rotation of the $|j_{\mathrm{eff}}=1/2,m \rangle$
state at each site $i$, thereby generating a spin-dependent hopping term.
In Sr$_{2}$IrO$_{4}$, where the IrO$_{6}$ octahedron is rotated by a angle
$\theta\approx 11^{\circ}$ about the $c$-axis, the effective hopping
matrix $\bar{t}^{ij}_{mm'}$ can be represented in terms of Pauli matrices
by $\mathsf{t}^{ij} = \bar{t}_{0} \mathbf{1} + i \bar{t}_{1} \sigma_{z}$
where $\bar{t}_{0}$ and $\bar{t}_{1}$ for $(ij)=\hat{x}$ or $\hat{y}$
become
\begin{equation}
  \label{eq:10}
  \bar{t}_{0}^{\hat{x}/\hat{y}} = \frac{2t_{0}}{3}\cos\theta
  (2\cos^{4}\theta - 1)
\end{equation}
\begin{equation}
  \label{eq:11}
  \bar{t}_{1}^{\hat{x}/\hat{y}} = \frac{2t_{0}}{3}
  \sin\theta(2\sin^{4}\theta - 1) .
\end{equation}

At half-filling, an effective $j_{\mathrm{eff}}$-spin Hamiltonian can be derived from the
$j_{\mathrm{eff}}$=1/2 single-band Hubbard model of Eq.~\eqref{eq:8}:
\begin{equation}
  \label{eq:12}
  \mathcal{H}_{\mathrm{spin}} = \sum_{\langle ij\rangle} \left[ I_{0}
  \mathbf{J}_{i}\cdot\mathbf{J}_{j} + I_{1}
  J_{z i}J_{z j} + \mathbf{D}_{ij}\cdot
  \mathbf{J}_{ i}\times\mathbf{J}_{j} \right]
\end{equation}
where $I_{0} = 4(\bar{t}_{0}^{2}-\bar{t}_{1}^{2})/\bar{U}$, $I_{1} =
8\bar{t}_{1}^{2}/\bar{U}$, and $\mathbf{D}_{ij} = D_{z}\hat{\mathbf{z}}$
with $D_{z}=8\bar{t}_{0}\bar{t}_{1}/\bar{U}$.
The first term is a conventional Heisenberg form of superexchange with the
coupling constant $I_{0}$. The second and third terms are pseudo-dipolar and
Dzyaloshinkii-Moriya (DM) antisymmetric exchange interactions, which originate
from  the pure imaginary hopping matrix element $i\bar{t}_{1}$ between the
neighboring $|j_{\mathrm{eff}}=1/2,m \rangle$ states of rotated IrO$_{6}$
octahedra.

\begin{figure}[t]
 \centering\includegraphics[width=8cm]{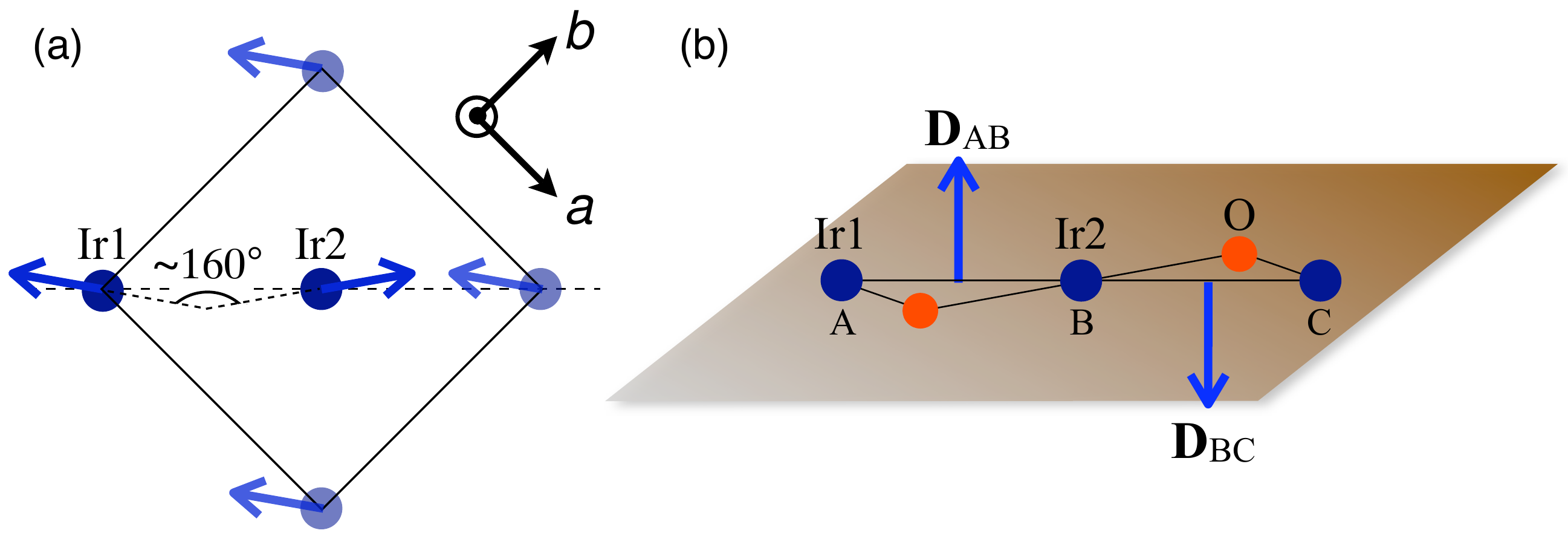}
 \caption{(Color online) (a) Magnetic configuration and (b) DM vectors of
   the calculated LDA+SO+$U$ ground state of Sr$_{2}$IrO$_{4}$. Blue
   arrows in (a) represent a non-collinear ordering of the local Ir
   moments, consisting of both spin and orbital components, in a canted
   AFM configuration.  The DM vectors, $\mathbf{D}_{\mathrm{AB}}$ and
   $\mathbf{D}_{\mathrm{BC}}$, in (b) are aligned along the $c$-axis with
   alternating signs and consistent with the Dzyaloshinkii-Moriya (DM)
   rule. } \label{fig:3}
\end{figure}

\subsection{Comparison with LDA+SO+$U$ Results}
\label{sec:anis-exch-inter-2}

From our LDA+SO+$U$ calculations, the magnetic configuration of the
insulating ground state was determined to be a canted antiferromagnetic
(AFM) state with the $ab$-plane as an easy plane.  We found no preferred
direction within the $ab$-plane.  As illustrated in Fig.~\ref{fig:3}(a),
there are two inequivalent Ir sites, i.e., Ir1 and Ir2 within the
$\sqrt{2}\times\sqrt{2}$ unit cell. It is found that the magnetic moment
at each Ir site is 0.36 $\mu_{\mathrm{B}}$ and both spin (0.10
$\mu_{\mathrm{B}}$) and orbital (0.26 $\mu_{\mathrm{B}}$) moments are
parallel to each other. In addition, AFM moments are canted with the
canting moment 0.063 $\mu_{\mathrm{B}}$, which is comparable to the single
crystal measurement \cite{PhysRevB.57.R11039}.

According to the rule by Dzyaloshinkii and Moriya \cite{yosida}, the
direction of the vector $\mathbf{D}_{ij}$ in Sr$_2$IrO$_4$ should point to
the $c$-axis due to a mirror plane containing Ir1-O-Ir2, as illustrated in
Fig.~\ref{fig:3}(b). The directions of $\mathbf{D}_{ij}$ can be
represented by $\mathbf{D}_{\mathrm{AB}} = - \mathbf{D}_{\mathrm{BC}} =
(0,0,d_{c})$, when considering the inversion symmetry at the site B, which
gives the consistent results as the $j_{\mathrm{eff}}$=1/2 Hamiltonian of
Eq.~\eqref{eq:12}. From the LDA+SO+$U$ calculations, it is concluded that
the DM interaction is responsible for the magnetic anisotropy of
Sr$_{2}$IrO$_{4}$ with the $ab$-plane as an easy plane but isotropic
within the $ab$-plane, whereas the single-ion anisotropy term has a
negligible contribution. Contrary to the
La$_{2}$CuO$_{4}$,\cite{cheong89:prb} which has no single-ion anisotropy
due to the $S$=1/2 ground state, the absence of the $ab$-plane anisotropy
in Sr$_{2}$IrO$_{4}$ is attributed to the tetragonal symmetry.

From the effective exchange Hamiltonian of Eq.~\eqref{eq:12}, the ratio of
$D_{z}/I_{0}$, which determines the spin canting angle, becomes
$|D_{z}/I_{0}| \approx \tan 2\theta$ for small $\theta$. In the strong SO
coupling limit, the canting angle increases close to the rotation angle of
IrO$_6$ octahedra.  We can estimate the magnitude of $\mathbf{D}_{ij}$ to
be $|\mathbf{D}| \approx 3.8$ meV assuming the intersite superexchange
interaction $J \approx 10$ meV. This enormous DM interaction may well be
related to the peculiar nature of the $j_{\mathrm{eff}}$=1/2 state.
Contrary to the $S$=1/2 counterpart of
La$_{2}$CuO$_{4}$,\cite{cheong89:prb} the $j_{\mathrm{eff}}$=1/2 state has
an open-shell of the $l=1$ orbital where the non-perturbative ground state
of $j_{\mathrm{eff}}$=1/2 spin-orbit coupled state contribute to the DM
term. Although the small magnetic moment of Ir observed in experiments was
attributed to the effective moment the $j_{\mathrm{eff}}$=1/2 state, one
can expect possible contributions from the $j$=1/2
quantum-fluctuation. Nevertheless, since $j_{\mathrm{eff}}$=1/2 state is
an eigenstate of the fictitious angular moment
$\mathbf{J}_{\mathrm{eff}}=\mathbf{L}_{\mathrm{eff}}+\mathbf{S}=-\mathbf{L}+\mathbf{S}$, 
the orbital contribution to the magnetic moment needs a careful interpretation \cite{kim:076402}.

\section{Conclusions}
\label{sec:conclusions}

In summary, we presented the effective $j_{\mathrm{eff}}$-spin model
Hamiltonian for Sr$_{2}$IrO$_{4}$. The strong SO interaction combined with
the large crystal field splitting in 5$d$ TMOs introduces a unique form of
the spin-orbit integrated band state at $E_{\mathrm{F}}$, leading to an
effective insulating ground state of $j_{\mathrm{eff}}$=1/2 quantum
magnet. The observed weak ferromagnetism is understood by the DM
anisotropic exchange interaction where the effective exchange interactions
arise from the full spin-orbital Hilbert space.  We hope that our
prototype model of the spin-orbit integrated magnetism is useful for the
study of various spin-orbit entangled physics. By taking an analogy of the
high $T_{c}$ superconductors as a doped $S$=1/2 quantum magnet, it will be
interesting to observe a doped $j$=1/2 quantum magnet as a spin-orbit
integrated correlated electron system.

\begin{acknowledgments}
  We are grateful to Profs.\ T.~W.~Noh and J.~H.~Park for valuable
  comments and suggestions. This work was supported by the KOSEF through
  the ARP (R17-2008-033-01000-0). We also acknowledge the computing
  resources support by the KISTI Supercomputing Center.
\end{acknowledgments}


\begin{thebibliography}{26}
\bibitem[{foo()}]{footnote}
\bibinfo{note}{We have tried an unphysically large $U$=6eV in LDA+$U$ without
  the SO coupling and then obtained an insulating ground state with a large
  spin and orbital polarization, which may lead to a lattice distortion.}
\end{thebibliography}
\end{document}